\documentclass{epl}

\newcommand{\ie}{{\it i.e.}~}
\newcommand{\eg}{{\it e.g.}}

\newcommand{\fig}{fig.}

\newcommand{\Ref}{ref.}
\newcommand{\Refs}{refs.}

\title{Could one find petroleum using neutrino oscillations in matter?}
\shorttitle{Could one find petroleum \ldots?}
\author{Tommy Ohlsson\thanks{E-mail: \email{tohlsson@ph.tum.de}} \and 
Walter Winter\thanks{E-mail: \email{wwinter@ph.tum.de}}}
\institute{
Institut f{\"u}r Theoretische Physik, Physik-Department,
Technische Universit{\"a}t M{\"u}nchen, James-Franck-Stra{\ss}e, 85748
Garching bei M{\"u}nchen, Germany
}

\begin{document}

\maketitle

\begin{abstract}
In neutrino physics, it is now widely believed that neutrino
oscillations are influenced by the presence of matter, modifying the
energy spectrum produced by a neutrino beam traversing the
Earth. Here, we will discuss the reverse problem, \ie what could be
learned about the Earth's interior from a single neutrino baseline
energy spectrum, especially about the Earth's mantle.
We will use a statistical analysis with a low-energy neutrino beam under very
optimistic assumptions. At the end, we will note that it is hard to find
petroleum with such a method, though it is not too far away from
technical feasibility.
\end{abstract}

Recently, neutrino physics and especially neutrino oscillations have
drawn a lot of attention in the field of physics. This is mainly due
to the successes of the Super-Kamiokande and SNO experiments
\cite{Fukuda:1998mi,Ahmad:2001an,Bahcall:2001ti}, which
have strongly indicated that neutrinos are massive particles and that they are
oscillating among different flavours. As far as we know today, the
neutrinos come in three flavours \cite{Groom:2000in}, \ie the
electron, muon, and tau neutrino. Neutrino oscillations are a purely
quantum mechanical phenomenon due to interference among the
flavours. In order to reveal further basic properties of
neutrinos and to pursue the mounting evidence for neutrino
oscillations, so-called neutrino factories have been proposed
\cite{Geer:1998iz,Albright:2000xi}. Exploiting some of the properties
of neutrinos, various approaches to {\it neutrino absorption tomography}, a
method in some sense similar to X-ray tomography, have been suggested
to obtain information on the interior of the Earth
\cite{Volkova74,DeRujula83,Wilson84,Askar84,Borisov87,Nicolaidis91,Crawford95,Kuo95,Jain:1999kp}.
However, these techniques face several difficulties involving
extremely high-energetic neutrino sources, large detectors, and the
prerequisite of many baselines. As a completely different approach,
the question has been raised if one could use the fact that neutrino
oscillations are influenced by the presence of matter
\cite{mikh85,mikh86,wolf78} to perform {\it neutrino oscillation tomography}
\cite{Ermilova:1988pw,Chechin:1991,Ohlsson:2001ck}, which would, in
principle, be possible with only one single baseline. However, movable
detectors, such as a grid of photomultipliers hanging from a movable
floating pontoon, have been proposed to be used together with an
upgraded CERN beam \cite{Dydak:2002}. In comparison to
geophysics, one could access the matter density profile directly with
the neutrino oscillation tomography method as
opposed to measuring the seismic wave velocity profile (see, \eg,
\Refs~\cite{aki80,lay95}). We will use a rather simple approach to see
what could be learned about cavities in the Earth's mantle from
neutrino oscillations in matter under very optimistic assumptions.

\begin{figure}
\begin{center}
\onefigure{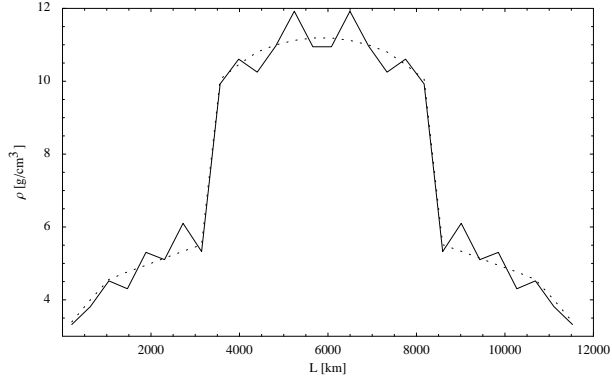}
\end{center}
\caption{The PREM (Preliminary Earth Reference Matter) density profile
\cite{stac77,dzie81} (dashed curve) as well as one possible
reconstructed profile from \Ref~\cite{Ohlsson:2001ck} close to the
$1 \sigma$ contour (solid curve) for a baseline length of $11{,}736
\, \mathrm{km}$ and the parameters used in
\Ref~\cite{Ohlsson:2001ck}. In this paper, the matter density
profile was cut into layers of constant matter densities to be
measured by a low-energy neutrino beam. Since the matter density of every
layer was treated as a free parameter and the contours of a
high-dimensional parameter space cannot be directly displayed, we only
show one representative of a possible matter density profile
close to the $1 \sigma$ contour to demonstrate the unability to
resolve short-scale fluctuations, \ie, one cannot exceed the
precision illustrated by this reconstructed matter density profile.}
\label{figure1}
\end{figure}

Thus, let us now assume a future scenario at the mid 21st century, when
neutrino factories have been operating already for some decades, and
the neutrino mixing parameters, such as mixing angles, phases, and mass
squared differences, have been measured more accurately. At this time,
we will probably know much more about neutrino oscillation technology
to build larger neutrino sources producing higher neutrino event
rates than proposed today. It was shown in \Ref~\cite{Ohlsson:2001ck}
that one can, in such a scenario, reconstruct the symmetric Earth's
matter density profile from a single neutrino baseline energy spectrum up to a
certain precision. Since the operators in the Hamiltonian 
describing neutrino propagation through different layers of matter are, in
general, non-commuting (see, \eg, \Ref~\cite{Ohlsson:1999um}),
a single baseline supplies more information on the matter density
profile than a single baseline in neutrino absorption tomography. In other
words, interference effects among the quantum mechanical transition amplitudes
of different density layers contain this information. However, one cannot
resolve density fluctuations of small amplitudes around the average value
of the matter density profile, as can be seen in \fig~\ref{figure1}. 
But what about cavities in the Earth's
mantle?
How large do they have to be in order to be identified in a
single neutrino baseline energy spectrum? All these questions points 
towards neutrino oscillation tomography, \ie using neutrino
oscillations in matter to learn something about the structure and
composition of the matter inside the Earth.

In order to have a neutrino beam very sensitive to matter density fluctuations
in the Earth's mantle, we use a low-energy neutrino beam
with about $500 \, \mathrm{MeV}$ for our investigation, as it is right now
often proposed for upgraded conventional beams (so-called superbeams)
\cite{Barger:2000nf,Itow:2001ee,Minakata:2001qm}.
In addition, we make very optimistic assumptions for cross sections, beam
characteristics, energy uncertainties, and detector properties, since
a possible experimental setup at this time can only be estimated. Let
us take $20$ energy bins between $300 \, \mathrm{MeV}$ and $500 \,
\mathrm{MeV}$ at a cross section proportional to $E^{1.66}$ in this
energy range \cite{Messier:1999kj}, where $E$ is the
energy\footnote{Note that for energies lower than about $1 \, 
  \mathrm{GeV}$ there are quite large uncertainties, which will need
  to be reduced by future experiments. However, it turns out that our
  application is rather insensitive to the slope of the cross section,
\ie the coefficient of the energy dependence.}, such that we could see
as many as an accumulated $10{,}000$ events per energy bin at $500 \,
\mathrm{MeV}$ to be folded with the neutrino oscillation transition
probabilities. This should be a reasonable guess without taking into
account too many yet unknown problem-specific details. For the
neutrino oscillations we choose the channel $\nu_{\mu} \rightarrow
\nu_e$ in a three-flavour neutrino oscillation analysis with the
parameter values $\theta_{13} = 5^\circ$, $\theta_{12} = \theta_{23} =
45^\circ$ (bimaximal mixing), $\Delta m^2_{21} = 3.65 \cdot 10^{-5} \,
\mathrm{eV}^2$, $\Delta m^2_{32} \simeq \Delta m^2_{31}= 2.5 \cdot
10^{-3} \, \mathrm{eV}^2$ and the CP phase $\delta =0$. (Of course,
the number of events in this channel depends on the value of the
mixing angle $\theta_{13}$.)

\begin{figure}
\begin{center}
\onefigure{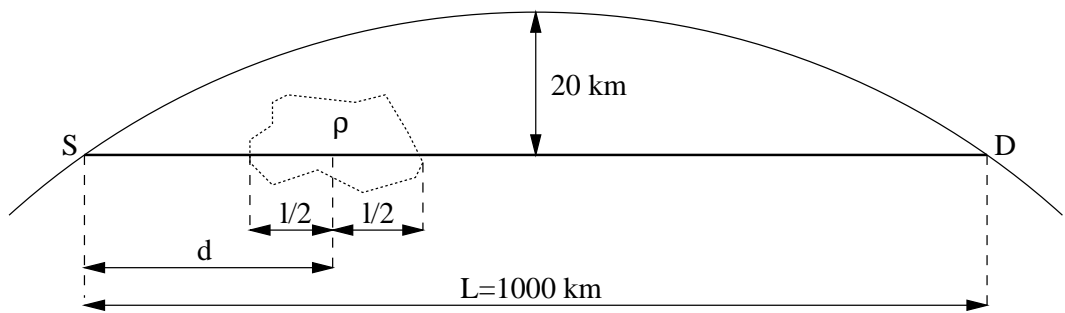}
\end{center}
\caption{A baseline configuration from a source $S$ to a
detector $D$ with a baseline length of $L=1{,}000 \, \mathrm{km}$, which
reaches a maximum depth of about $20 \, \mathrm{km}$ under the Earth's
surface, but has an average depth of about $13 \,\mathrm{km}$. The
neutrinos are propagating in matter of constant density of about $2.9
\, \mathrm{g/cm^3}$ and crossing a cavity of length $l$ centred at a
distance $d$ from the source. The matter density in the cavity is $\rho$.}
\label{figure2}
\end{figure}

Let us now look at a beam configuration as it is
shown in \fig~\ref{figure2}.
In this configuration, the neutrino beam, propagating in approximately
constant matter density in the Earth's mantle, crosses a cavity with matter
density $\rho$ and length $l$ centred at a distance $d$ from the
source $S$. We will speak about $l$ and $d$ as the size and position
of the cavity, respectively.
Assuming a neutrino energy spectrum measured at the detector $D$ produced with
the parameter values above, what can we learn about the parameters
$d$ and $l$? Since the phase shift in neutrino oscillations will depend on the
matter density contrast between the surrounding matter and the cavity, we
assume a rather small matter density within the cavity, \ie
$\rho \simeq 1 \, \mathrm{g/cm^3}$, corresponding to a cavity filled with
water.
The matter density contrast would be much larger for air-filled
cavities and much smaller for a porous rock, which may act as a petroleum trap.

\begin{figure}
\begin{center}
\onefigure{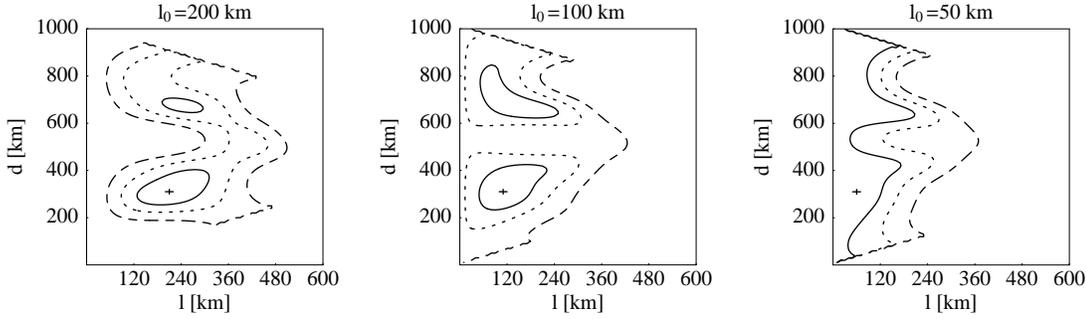}
\end{center}
\caption{The $1 \sigma$ (solid curve), $2 \sigma$ (dotted
curve) and $3 \sigma$ (dashed curve) contours in the statistical analysis of
a true cavity centred at $d_0=300 \, \mathrm{km}$. In each plot, the value
$l_0$ for the true cavity length is given in the title as well as the true
position is marked with a cross.}
\label{figure3}
\end{figure}

Figure~\ref{figure3} shows the results of a two-parameter statistical
analysis of a cavity centred at $d_0=300 \, \mathrm{km}$ for different
values of $l_0 \in \{50,100,200\} \, \mathrm{km}$.
In these plots, the true position $d_0$ and size $l_0$ of the cavity assumed
or measured is marked with a cross. The
contour lines tell us for the fixed reference values that we will
measure a value within the $1 \sigma$ contour with $68.3 \%$, within the $2
\sigma$ contour with $95.5 \%$ and within the $3 \sigma$ contour with $99.7
\%$ probability. Hence, these contours help us to estimate the measurability of
the parameters $d$ and $l$. Closed contours mean that we can really detect this
cavity within the configuration assumed on a statistical basis on the
respective significance level.
Contours open to the left only constrain the values within these, \ie
for a certain position $d$, cavities larger than a certain size $l$ can be
excluded from the measured neutrino energy spectrum. Small enough
cavities, however, could be located anywhere. 

In the plots in \fig~\ref{figure3}, large cavities with $l \simeq
200 \, \mathrm{km}$ can be clearly detected on the $3 \sigma$ level,
though with some degeneracy in position and uncertainty in size. The
degeneracy in the position $d$ basically comes from the periodicity in
neutrino oscillations and this may be resolved by using additional
knowledge coming from geophysics. Note that two-flavour neutrino
oscillations cannot distinguish time-inverted matter density profiles
\cite{Akhmedov:2001kd}, which means that the symmetry in the plots
with respect to the ``$d=500 \, \mathrm{km}$ line'' is only destroyed
by three-flavour effects. Furthermore, the areas fulfilling 
$d-l/2 < 0$ and $d+l/2 > 1{,}000 \, \mathrm{km}$ are excluded by
definition. Smaller cavities, such as for $l \simeq 100 \, \mathrm{km}$, can
only be seen on the $1 \sigma$ level, or even not at all, such as for $l
\simeq 50 \, \mathrm{km}$. The reason for this is the short cavity length
compared with the characteristic length scale of neutrino
oscillations, which is of the order of $1{,}000 \, \mathrm{km}$, as well
as the quite small matter density contrast.
Generally speaking, in any quantum mechanical problem, the influence of a
perturbing potential depends on its integral, \ie the length scale of the
perturbation times its amplitude.

\begin{figure}
\begin{center}
\onefigure{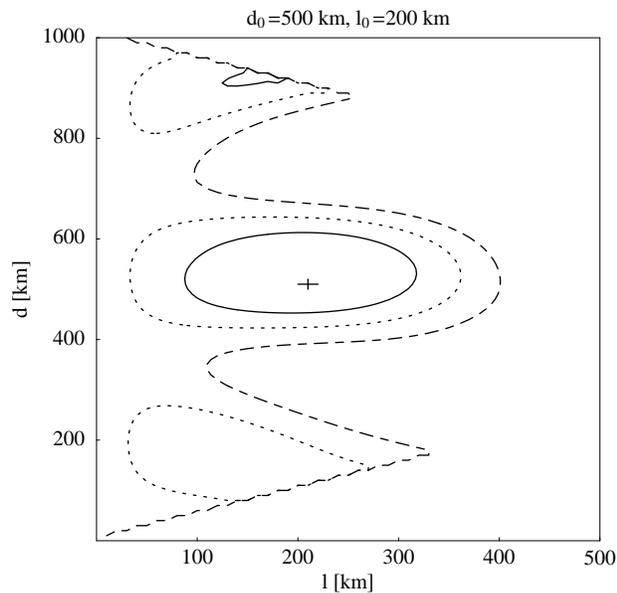}
\end{center}
\caption{The $1 \sigma$ (solid curve), $2 \sigma$ (dotted
curve) and $3 \sigma$ (dashed curve) contours in the statistical analysis
of a true cavity with $d_0=500 \, \mathrm{km}$ and $l_0 = 200 \, \mathrm{km}$.
}
\label{figure4}
\end{figure}

The result of the analysis is, however, not only dependent on the cavity size
$l_0$, but also on the cavity position $d_0$. Figure~\ref{figure4}
shows the result of a calculation with $d_0=500 \, \mathrm{km}$ and
$l_0 = 200 \, \mathrm{km}$, \ie the cavity is situated on the very
centre of the baseline.
In this case, the degeneracy in the position on the $1 \sigma$ level is three
compared with two above. In addition, the cavity can clearly be located on
the $2 \sigma$ level. However, on the $3 \, \sigma$ level, it cannot be
proven to exist at all. Thus, the position of the
cavity slightly modifies the statistics and has to be taken into account.

Coming back to our original idea, the search for petroleum, what can we
conclude from the above analysis? Because we used very
optimistic estimates for the source and detector, as well as we neglected
energy uncertainties, backgrounds, and cross section and mixing parameter
uncertainties, it may be very hard to exploit this idea. Moreover,
the cavity, \ie the porous rock acting as petroleum
trap, has to be quite large and it has to have a rather large density
contrast to the surrounding matter in order to be detected.
One can easily imagine that reducing the density contrast or taking
into account backgrounds and other uncertainties would make the closed
contours vanish, \ie the corresponding cavity could not be detected
anymore. However, with this sort of analysis one
should be able to see larger scale structures, such as cavity
systems, though conventional geophysical methods could be more
successful and less expensive. 
Thus, neutrino oscillation tomography may not be the very best
way to look for cavities in the Earth's mantle; however, it could, in
the far future, have different applications such as for the Earth's
core or other planets.
Finally, as a curiosity, taking the recent development and progress in
neutrino physics into account, it is interesting to observe that
neutrino oscillation tomography is not as far away from technical
feasibility as one may expect.

\acknowledgments
This work was supported by the Swedish Foundation for International
Cooperation in Research and Higher Education (STINT) [T.O.], the Wenner-Gren
Foundations [T.O.], the ``Studienstiftung des deutschen Volkes'' (German
National Merit Foundation) [W.W.], and the ``Sonderforschungsbereich 375
f{\"u}r Astro-Teilchenphysik der Deutschen Forschungsgemeinschaft''.


\begin{thebibliography}{0}

\bibitem{Fukuda:1998mi}
  \Name{Fukuda, Y. \etal, Super-Kamiokande Collaboration}
  \REVIEW{Phys. Rev. Lett.}{81}{1998}{1562}, arXiv:hep-ex/9807003.

\bibitem{Ahmad:2001an}
  \Name{Ahmad, Q. R. \etal, SNO Collaboration}
  \REVIEW{Phys. Rev. Lett.}{87}{2001}{071301}, arXiv:nucl-ex/0106015.

\bibitem{Bahcall:2001ti}
  \Name{Bahcall, J. N.}
  \REVIEW{Nature}{412}{2001}{29}.

\bibitem{Groom:2000in}
  \Name{Groom, D. E. {\it et al.}, Particle Data Group}
  \REVIEW{Eur. Phys. J. C}{15}{2000}{1}, http://pdg.lbl.gov/.

\bibitem{Geer:1998iz}
  \Name{Geer, S.}
  \REVIEW{Phys. Rev. D}{57}{1998}{6989},
  \SAME{59}{1999}{039903(E)}, arXiv:hep-ph/9712290.

\bibitem{Albright:2000xi}
  \Name{Albright, C. \etal} arXiv:hep-ex/0008064.

\bibitem{Volkova74}
  \Name{Volkova, L. V. \and Zatsepin, G. T.}
  \REVIEW{Bull. Phys. Ser.}{38}{1974}{151}.

\bibitem{DeRujula83}
  \Name{De R{\'u}jula, A., Glashow, S. L., Wilson, R. R. \and Charpak,
  G.}
  \REVIEW{Phys. Rep.}{99}{1983}{341}.

\bibitem{Wilson84}
  \Name{Wilson, T. L.}
  \REVIEW{Nature}{309}{1984}{38}.

\bibitem{Askar84}
  \Name{Askar'yan, G. A.}
  \REVIEW{Usp. Fiz. Nauk}{144}{1984}{523},
  [\REVIEW{Sov. Phys. Usp.}{27}{1984}{896}].

\bibitem{Borisov87}
  \Name{Borisov, A. B., Dolgoshein, B. A. \and Kalinovski\u{\i},
  A. N.}
  \REVIEW{Yad. Fiz.}{44}{1986}{681},
  [\REVIEW{Sov. J. Nucl. Phys.}{44}{1987}{442}].

\bibitem{Nicolaidis91}
  \Name{Nicolaidis, A., Jannane, M. \and Tarantola, A.}
  \REVIEW{J. Geophys. Res.}{96}{1991}{21811}.

\bibitem{Crawford95}
  \Name{Crawford, H. J., Jeanloz, R., Romanowicz, B. \and the
  DUMAND~Collaboration}
  Proc. of the XXIVth International Cosmic Ray Conference
  (University of Rome, Rome) 1995, p. 804.

\bibitem{Kuo95}
  \Name{Kuo, C. \etal}
  \REVIEW{Earth Plan. Sci. Lett.}{133}{1995}{95}.

\bibitem{Jain:1999kp}
  \Name{Jain, P., Ralston, J. P. \and Frichter, G. M.}
  \REVIEW{Astropart. Phys.}{12}{1999}{193}, arXiv:hep-ph/9902206.

\bibitem{mikh85}
  \Name{Mikheyev, S. P. \and Smirnov, A. Yu.}
  \REVIEW{Yad. Fiz.}{42}{1985}{1441},
  [\REVIEW{Sov. J. Nucl. Phys.}{42}{1985}{913}].

\bibitem{mikh86}
  \Name{Mikheyev, S. P. \and Smirnov, A. Yu.}
  \REVIEW{Nuovo Cimento C}{9}{1986}{17}.

\bibitem{wolf78}
  \Name{Wolfenstein, L.}
  \REVIEW{Phys. Rev. D}{17}{1978}{2369}.

\bibitem{Ermilova:1988pw}
  \Name{Ermilova, V. K., Tsarev, V. A. \and Chechin, V. A.}
  \REVIEW{Bull. Lebedev Phys. Inst.}{NO.3}{1988}{51}.

\bibitem{Chechin:1991}
  \Name{Chechin, V. A. \and Ermilova, V. K.}
  Proc. of LEWI'90 School (Dubna) 1991, p. 75.

\bibitem{Ohlsson:2001ck}
  \Name{Ohlsson, T. \and Winter, W.}
  \REVIEW{Phys. Lett. B}{512}{2001}{357}, arXiv:hep-ph/0105293.

\bibitem{Dydak:2002}
  \Name{Dydak, F.}
  talk given at the XXth International Conference on Neutrino Physics
  \& Astrophysics (Neutrino 2002), Munich, Germany, 2002.

\bibitem{aki80}
  \Name{Aki, K. \and Richards, P. G.}
  \Book{Quantitative Seismology - Theory and Methods}
  \Vol{1, 2}
  \Publ{W. H. Freeman, San Francisco}
  \Year{1980}.

\bibitem{lay95}
  \Name{Lay, T. \and Wallace, T. C.}
  \Book{Modern Global Seismology}
  \Publ{Academic Press, New York}
  \Year{1995}.

\bibitem{Ohlsson:1999um}
  \Name{Ohlsson, T. \and Snellman, H.}
  \REVIEW{Phys. Lett. B}{474}{2000}{153},
  \SAME{480}{2000}{419(E)}, arXiv:hep-ph/9912295.

\bibitem{stac77}
  \Name{Stacey, F. D.}
  \Book{Physics of the Earth}
  \Publ{Wiley, New York}
  \Year{1977}, 2nd ed.

\bibitem{dzie81}
  \Name{Dziewonski, A. M. \and Anderson, D. L.}
  \REVIEW{Phys. Earth Planet. Inter.}{25}{1981}{297}.

\bibitem{Barger:2000nf}
  \Name{Barger, V., Geer, S., Raja, R. \and Whisnant, K.}
  \REVIEW{Phys. Rev. D}{63}{2001}{113011}, arXiv:hep-ph/0012017.

\bibitem{Itow:2001ee}
  \Name{Itow, Y. \etal} arXiv:hep-ex/0106019.

\bibitem{Minakata:2001qm}
  \Name{Minakata, H. \and Nunokawa, H.}
  \REVIEW{J. High Energy Phys.}{10}{2001}{001}, arXiv:hep-ph/0108085.

\bibitem{Messier:1999kj}
  \Name{Messier, M. D.} Evidence for neutrino mass from observations of
  atmospheric neutrinos with Super-Kamiokande, Ph.D. thesis,
  UMI-99-23965, Boston University, 1999.

\bibitem{Akhmedov:2001kd}
  \Name{Akhmedov, E., Huber, P., Lindner, M. \and Ohlsson, T.}
  \REVIEW{Nucl. Phys. B}{608}{2001}{394}, arXiv:hep-ph/0105029.

\end{thebibliography}
\end{document}